\documentclass[12pt]{article}
\usepackage{graphicx,amsmath}
\usepackage{units}
\usepackage{color}

\newlength{\dinwidth}
\newlength{\dinmargin}
\renewcommand{\vec}[1]{\boldsymbol{#1}}

\def\lapproxeq{\lower .7ex\hbox{$\;\stackrel{\textstyle                                                    
<}{\sim}\;$}}                                                    
\def\gapproxeq{\lower .7ex\hbox{$\;\stackrel{\textstyle                                                    
>}{\sim}\;$}}                                                    
\def\be{\begin{equation}}                                                    
\def\ee{\end{equation}}                                                    
\def\bea{\begin{eqnarray}}                                                    
\def\eea{\end{eqnarray}}
\def\b{\vec{b}}
     
\def\q{\vec{q}}

\def\GeV{\rm GeV}

\def\sh{\hat s}
\def\sh2{{\hat s}^2}

\begin{document}
\titlepage                                                    
\begin{flushright}                                                    
IPPP/20/65  \\                                                    
\today \\                                                    
\end{flushright} 
\vspace*{0.5cm}
\begin{center}                                                    
{\Large \bf Dynamics of diffractive dissociation}\\
\vspace*{1cm}
                                                   
V.A. Khoze$^{a,b}$, A.D. Martin$^a$ and M.G. Ryskin$^{a,b}$ \\                                                    
                                                   
\vspace*{0.5cm}                                                    
$^a$ Institute for Particle Physics Phenomenology, University of Durham, Durham, DH1 3LE \\                                                   
$^b$ Petersburg Nuclear Physics Institute, NRC Kurchatov Institute, Gatchina, St.~Petersburg, 188300, Russia

\vspace*{1cm}

\begin{abstract}

  We describe a QCD based model which incorporates the main properties of  
  the inclusive particle distributions expected for diffractive 
  processes, including the diffractive dissociation at high energies.
  We study, in turn, the total cross section, $\sigma_{\rm tot}$, the differential elastic, 
  $d\sigma_{\rm el}/dt$, cross section, the dependence of the single proton dissociation cross section,
$\xi d\sigma^{\rm SD}/d\xi$, on the  momentum fraction, $\xi=1-x_L$, lost  
by the leading proton, the multiplicity distributions in 
inelastic (non-diffractive) collisions and in the processes of 
dissociation. Besides this we calculate the mean 
transverse momenta of the `wee partons' (secondaries) produced  in the case of dissociation (that is in the
processes with a large rapidity gap) and compare it with that in inelastic interactions.  
  
\end{abstract}
\end{center}  

\vspace*{1cm}

\section{Introduction}

In a recent paper~\cite{Luk} the inclusive distribution of identified 
particles produced in Single Diffractive Dissociation (SD) $pp\to p+X$ 
processes were studied with the STAR detector at RHIC in
  proton-proton collisions at $\sqrt s=200$ GeV. Here $X$ denotes the 
  diffractively produced system. The SD  events were selected by 
  observing in the Roman Pot system(s) the
 leading proton (or protons) which carry a large fraction, $x_L$, of 
 the beam momentum. We denote  
$x_L=1-\xi$. Analogous experiments are underway or being planned by  
CMS-TOTEM (PPS) and ATLAS-AFP at the LHC. The leading proton is 
observed in the TOTEM or ALFA Roman Pots while the diffracted system $X$ 
is studied by the central CMS or ATLAS detectors (see e.g.~\cite{CT,AA}).

\begin{figure} [t]
\begin{center}
\includegraphics[trim=0.0cm 0cm 0cm 0cm,scale=0.7]{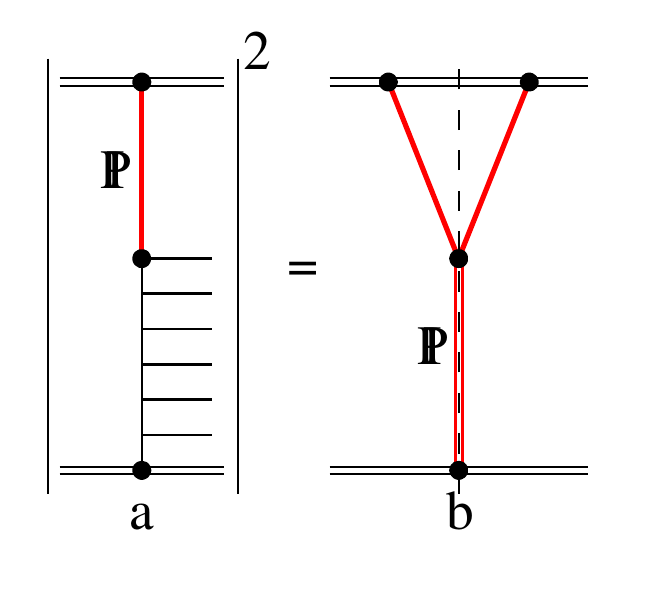}
\caption{\sf Schematic diagrams of Single Diffractive (SD) 
processes  in Pomeron-proton collisions.} 
\label{f1}
\end{center}
\end{figure}

Note that after the leading proton(s) with large $x_L$ close to 1 
are detected we have rather small remaining energy to produce the new 
secondaries. Therefore,
these new secondaries (system $X$) are separated from the leading 
proton(s) by Large Rapidity Gap(s) (LRG) with size\footnote{For 
$pp\to p+X$ the mass $M_X$ of the produced system $X$ is given by 
$M^2_X=s(1-x_L)$ with gap sizes $\Delta y\simeq -\ln(1-x_L)$.} 
$\Delta y\simeq$ ln$(1/\xi)$. Since the interaction across the LRG is
provided by the Pomeron exchange
such events can be interpreted as the result of a Pomeron interaction 
with a proton (SD). The processes are illustrated in Fig.~\ref{f1}a.

In the first approximation at large mass, $M_X$, of the system $X$ the Pomeron-proton interaction is driven by another Pomeron exchange and the cross section of whole process is described by the triple-Pomeron diagram Fig.\ref{f1}b.

However actually the situation is more complicated and the simple triple-Pomeron  diagram can be used only in the situation when the probability of interaction is relatively small and the parton densities are rather low. On another hand diffractive dissociation is a soft process and here we deal with {\em strong}  interactions. Therefore we have to consider the possibility of a few simultaneous interactions. Indeed, there is a rather large probability that 
 the LRG will be filled by secondaries produced in an additional soft interaction~\footnote{In Monte Carlo Models this is called the Multiple Parton Interaction option (MPI). Recall that in the present paper we consider only an {\em individual} $pp$-collision and do not account for the possibility of other proton-proton interactions in the same bunch crossing which occur if the instantaneous luminosity is large.} and instead of single proton dissociation (SD) we will observe a completely inelastic event. That is we have to account for the gap survival probability, $S^2<1$, which in terms of the Reggon Field Theory~\cite{G1968} is  
 described by the multi-Pomeron diagrams responsible for the absorptive corrections. For this reason the distributions of particles produced in non-diffractive inelastic collisions and in the processes with LRG become different.

  At a qualitative level the corresponding difference was discussed in~\cite{Pom}. In the present paper we consider
  a model which allows us to evaluate the expected difference (semi)quantitatively. We attempt to make our model relatively simple, but to keep the main properties of the development of the `wee parton' cascade \footnote{We use the term 'wee parton' in the spirit  of R. Feynman and V.N. Gribov~\cite{Fe,Gr} as some elementary object which participates in strong interactions and carries a very small part of the parent hadron momentum. In the case of QCD it is mainly the gluon. However in our simplified model we do not fix the quantum numbers of these partons.} ; namely we account for the `diffusion' in impact parameter, $b$, space and for the growth of the characteristic transverse momenta, $k_t$, when the parton density becomes large.
   These are the important features of the perturbative QCD evolution observed within the BFKL~\cite{BFKL} and multi-Pomeron approach (see e.g.~\cite{GLR,LR} and for a recent review~\cite {PDG}).

Recall that within perturbative QCD, the BFKL equation describes the rapidity ($y$) evolution of parton/gluon density (i.e. the proton opacity, $\Omega(y)$) and predicts the exponential growth $\Omega(y)\propto \exp(\omega_{{\rm BFKL}}~y)$. 
Moreover, at each step of the evolution the parton transverse momentum, $k_t$, may be changed few times in one or another direction and the position of a parton in impact parameter space, $b$, can be moved by $\Delta b\sim 1/k_t$. That is we have diffusion in both the $b$ and $\ln k_t$ spaces. The absorptive corrections make this diffusion asymmetric. Due to a larger parton density, and correspondingly to stronger absorptive corrections, in the centre of disk the partons are mainly moving in the direction of the periphery;  while the remaining partons occupy the larger $k_t$ elements of the  $(b,k_t)$ configuration space (see sect.3.2 of ~\cite{LR} for more details).

Looking for events with a LRG we select interactions occuring in the periphery of the disk where the probability of gap survival is larger. Thus in order to  reproduce the main feature of diffractive dissociation at high energies our model must include
\begin{itemize}
\item the growth of parton densities,
\item the possibility of movement in $b$-plane,
\item absorptive corrections during the $y$-evolution process, 
\item gap survival probabilities with respect to the rescattering  of the partons which belong to the different
(beam and target) incoming protons. 
\end{itemize}

In the next section we describe the structure of the evolution of  the `wee-parton' cascade. Then in sect.3 we present the formulae to calculate the total, elastic and diffractive dissociation cross sections and the multiplicity distributions of secondaries based on the resulting cascades. Numerical values of parameters used in the model are presented in sect.4, while in sect.5 we show the results obtained for SD processes. These results will be discussed in sect.6. 
We conclude in sect.7.

\section{Parton evolution}
Describing the evolution of the wee-parton cascade we will account for the absorptive effects caused by the possibility of an additional interaction between the parton and the parent proton~\footnote{We ignore here the parton-parton rescatterings. These give a smaller effect.}. That is, our approach includes not only the multiple interactions between the beam and target hadrons (protons) but also the multiple interactions between the particular parton and the proton as well. For this purpose we use the eikonal model. That is, we assume an eikonal-like form of the multi-Pomeron vertices. Specificly the coupling of $n$ to $m$ Pomerons takes the form 
\be
g^n_m=(g_N\lambda)^{n+m-2}\ ,
\ee
                                                                                                                                                                                                                                                                                                                                                                                                                                                                                                                                                                                                                                                                                                                                                                                                                                                                                                                                                                                                                                                                                                                                                                                                                                                                              where $g_N$ is the proton-Pomeron coupling and $\lambda$ accounts for the suppression of the triple-Pomeron vertex ($g_{3P}=g^1_2$) in comparison with $g_N$.

\subsection{Good-Walker formalism}
In the simplest case we have a one-channel eikonal model in which at hadron level we consider only elastic (intact proton) intermediate states.  To allow for the possibility of low mass $p\to N^*$ excitations (in the intermediate state), we need a multi-channel eikonal with $g_{pN^*}$ and $g_{N_a^*N_b^*}$ transition vertices. For this we use the Good-Walker  (G-W) formalism~\cite{GW} which  
introduces states $\phi_k$ that diagonalize  
the $T$ matrix of the high energy hadrons couplings (e.g. in the proton case describes different $p\to N^*,~ N^*_a\to N^*_b$ transitions).
Such eigenstates only undergo elastic
scattering since there are no off-diagonal transitions. That is
\be \langle \phi_i|T|\phi_k\rangle = 0\qquad{\rm for}\ i\neq k \ee
and so a state $k$ cannot diffractively dissociate in a state $i$.
Thus, working in terms of G-W eigenstates $\phi_i$,
we have a simple one-channel eikonal for each state.
We denote the orthogonal matrix which diagonalizes ${\rm
Im}\,T$ by $a$, so that
\be \label{eq:a3} {\rm Im}\,T \; = \; aF(a)^T \quad\quad {\rm with}
\quad\quad \langle \phi_i |F| \phi_k \rangle \; = \; F_k \:
\delta_{ik}, \ee
where $F_k$ is the probability of  the hadronic process proceeding via the
 diffractive eigenstate $\phi_k$.

Now consider the diffractive dissociation of an  incoming state
state $|i\rangle$. We can write
\be \label{eq:a4} | i \rangle \; = \; \sum_k \: a_{ik} \: | \phi_k
\rangle. \ee
The elastic scattering amplitude satisfies
\be \label{eq:a5} \langle i |{\rm Im}~T| i \rangle \; = \; \sum_k
\: |a_{ik}|^2 \: F_k \; = \; \langle F \rangle, \ee
where $F_k \equiv \langle \phi_k |F| \phi_k \rangle$. Here the
brackets of $\langle F \rangle$ mean that we take the average of
$F$ over the initial probability distribution of diffractive
eigenstates. After diffractive scattering described by
$G_{fi}$, the final state $| f \rangle$ will, in general, be a
different superposition of eigenstates from that the initial state $| i \rangle$,
which was shown in~(\ref{eq:a4}). Neglecting 
the real parts of the amplitudes for the moment, the cross
sections at a given impact parameter $b$, will have the forms
\bea \frac{d \sigma_{\rm tot}}{d^2 b} & = & 2 \:
{\rm Im} \langle i |T| i \rangle \; = \; 2 \: \sum_k
\: |a_{ik}|^2 \: F_k \; = \; 2 \langle F \rangle \label{eq:b1} \\
& & \nonumber\\
\frac{d \sigma_{\rm el}}{d^2 b} & = & \left | \langle i |T| i
\rangle \right |^2 \; = \; \left (
\sum_k \: |a_{ik}|^2 \: F_k \right )^2 \; = \; \langle F \rangle^2 \label{eq:b2}\\
& & \nonumber \\
\frac{d \sigma_{\rm el \: + \: SD}}{d^2 b} & = & \sum_k \: \left |
\langle \phi_k |T| \phi_i \rangle \right |^2 \; = \; \sum_k \:
|a_{ik}|^2 \: F_k^2 \; = \; \langle F^2 \rangle. \label{eq:b3} \eea
It follows that the cross section for the single diffractive
dissociation of a proton,
\be \label{eq:a7} \frac{d \sigma_{\rm SD}}{d^2 b} \; = \; \langle
F^2 \rangle \: - \: \langle F \rangle^2, \ee
is given by the statistical dispersion in the absorption
probabilities of the diffractive eigenstates. Here the average is
taken over the components $k$ of the incoming proton which
dissociates.

One consequence is the important result that if all the components $\phi_k$ of the incoming
proton $| i \rangle$ were absorbed equally then the
diffracted superposition would be proportional to the incident one
and  the inelastic diffraction would be zero.  Thus if, at very
high energies, the amplitudes $F_k$ at small impact parameters are
equal to the black disk limit, $F_k = 1$, then diffractive
production will be equal to zero in this impact parameter domain,
and so the dissociation will only occur in the peripheral $b$ region where the  edge of the disk is not 
   completely black.
 Hence the impact parameter
structure of diffractive dissociation and elastic scattering are drastically
different in the presence of absorptive $s$-channel unitarity effects.

In our simple model to account for the low mass dissociation~\footnote{The high mass, $M_X$, dissociation will be described in sect.3.} we will include two eigenstates which at the beginning of evolution ($y=0$) have different sizes, that is different impact parameter, $b$ distributions of the parton densities, but almost equal densities at $b=0$. To minimize the number of free parameters these two eeigenstates are taken with  equal probabilities, that is $a_{p1}=a_{p2}=1/\sqrt 2$.

\subsection{Rapidity evolution}
\begin{figure} [t]
\begin{center}
\includegraphics[trim=0.0cm 0cm 0cm 0cm,scale=0.7]{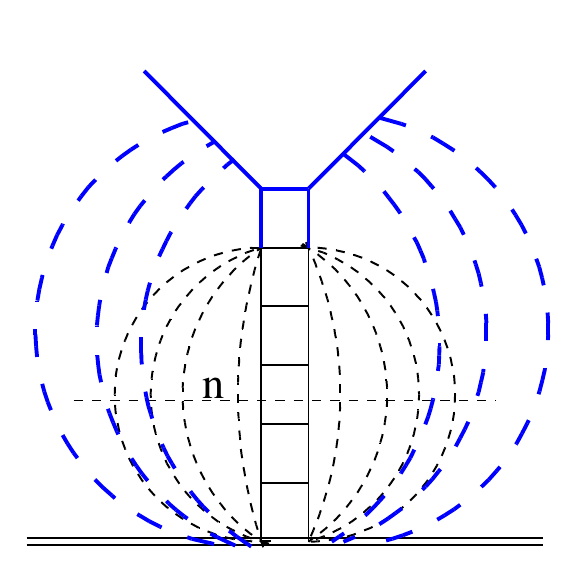}
\caption{\sf Evolution of the wee parton density in rapidity (momentum  fraction) space. The last step of evolution is shown by thick (blue) lines. The dashed curves indicate the eikonal-like absorptive corrections. $n$ is the number of screening Pomerons.}
\label{f2}
\end{center}
\end{figure}

Here we consider the evolution in rapidity $y$ of the parton cascade generated by an individual G-W eigenstate. 
To describe the evolution of the wee parton density in rapidity space we first neglect the diffusion in impact parameter, $b$, and omit the absorptive effects. For a fixed $b$ the optical density (opacity), $\Omega(b,y)$ evolves with decreasing momentum fraction, $x$, carried by the parton as
\begin{equation}
\label{e1}
\frac{d\Omega(b,y)}{dy}~=~\Delta\Omega(b,y)\quad\quad y=\ln(1/x) \, ,
\end{equation}
where  we expect that the value of $\Delta$ to be close to that ($\omega_{\rm BFKL}$) given by the BFKL~\cite{BFKL} intercept $\omega_{\rm BFKL}$. That is accounting for (and re-summing) the  next-to-leading logarithm corrections we expect $\Delta\sim 0.15-0.25$~\cite{NLL}. This evolution is indicated by the  continuous lines in Fig.\ref{f2}, where the thick (blue) lines  indicate the last step of the evolution in $y$.

At each step of the evolution we have to account for the absorptive effects caused by additional parton-target interactions which are shown in Fig.\ref{f2} by dashed curves. It is convenient now to deal  with the probability of an inelastic interaction $G$ rather than with the opacity $\Omega$ 
\begin{equation}
\label{e2}
G(b,y)~=~1-e^{-\Omega(b,y)} \, .
\end{equation}
Here we assume an eikonal form of the multi-Pomeron vertices. 
 Each step of the evolution is now suppressed by the survival factor $\exp(-\Omega)=1-G(b,y)$ and 
the evolution reads
\begin{equation}
\label{e3}
\frac{dG(b,y)}{dy}~=~\Delta (1-G(b,y))~G(b,y)\, .
\end{equation}
The factor $(1-G(b,y))$ provides the saturation of the parton density at $G\to 1$.

Next we have to include the diffusion in the transverse $b$-plane~\footnote{The diffusion in $b$ space was considered long ago in~\cite{FC,Grb}.}. This is an important effect which leads to the shrinkage of the diffractive cone (i.e. to the growth of the elastic $t$-slope, $B_{\rm el}$, with energy). 
At each step of the evolution the parton can move in $b$ space by some interval $\delta b\simeq 1/k_t$, where $k_t$ is the parton transverse momentum.

Actually the main effect is observed when the parton moves {\em outwards} from the centre of disk. Only this will be accounted for in our simplified model. We assume that one quarter (one of the four  (+x, -x, +y, -y) possible transverse directions) of the partons
generated at each step of the evolution goes to a larger value of $b$ with the probability
\begin{equation}
\label{e4}
\frac{dP(b)}{db}~=~k_t(b)\exp(-bk_t(b))\, ,
\end{equation}
where we consider only the movement outside of the centre of the disk and  $k_t(b)$ is the typical transverse momentum of the parton placed at impact parameter $b$. That is finally we obtain the evolution equation 
\begin{equation}
\label{emast}
\frac{dG(b,y)}{dy}~=~(1-G(b,y))\left[\frac 34\Delta G(b,y)+\frac 14\Delta \int_0^b  db'G(b',y)k_t(b',y)e^{(b'-b)k_t(b',y)}\right]\, .
\end{equation}
This should be complemented by the equation for $k_t(b,y)$.
 As far as the parton density approaches its saturation limit
  ($G\to 1$) the new partons start to occupy a larger $k_t$ region. Asymptotically we have to keep the probability of an additional interaction 
  $w=\sigma^{\rm abs}/\pi R^2=const$. Here $\pi R^2$ is the ``hot spot" area occupied by the parton cascade. In the first  approximation the absorptive cross section $\sigma^{\rm abs}$ increases with $y=\ln(1/x)$ as $\sigma^{\rm abs}\propto  (1/k^2_t)\exp(y\Delta)$. That is the transverse momentum $k_t$ grows as $k_t\propto\exp(y\Delta/2)$. Being far from the saturation limit we expect more or less constant $k_t$ but when the density $G\to 1$ approches saturation the value of $k_t$ starts to grow.
  Therefore we choose
\begin{equation}
\label{ekt}
\frac{dk_t(b,y)}{dy}~=~\frac{\Delta}2k_t(b,y)G(b,y) \, .
\end{equation}
These two equations (\ref{emast}) and (\ref{ekt}) describe
our simplified evolution of the wee parton cascade. When the parton density is small ($G\ll 1$) the new partons created at the current step of evolution 
have more or less the same $k_t$ as the parent parton and mainly enlarge the value of $G(b)$ at the same $b$ point, partly moving to the periphery of disk; that  is, to larger $b$. At a larger density $G$ this process is suppressed by the $1-G(b)$ factor. The `remaining' partons
 (see the last factor $G$ in (\ref{ekt})) start to occupy a larger $k_t$ space (see~\cite{LR} for more detailed description
 of the parton cascade development).

\begin{figure} [t]
\begin{center}
\includegraphics[trim=-3.0cm 0cm -3cm 0cm,scale=0.6]{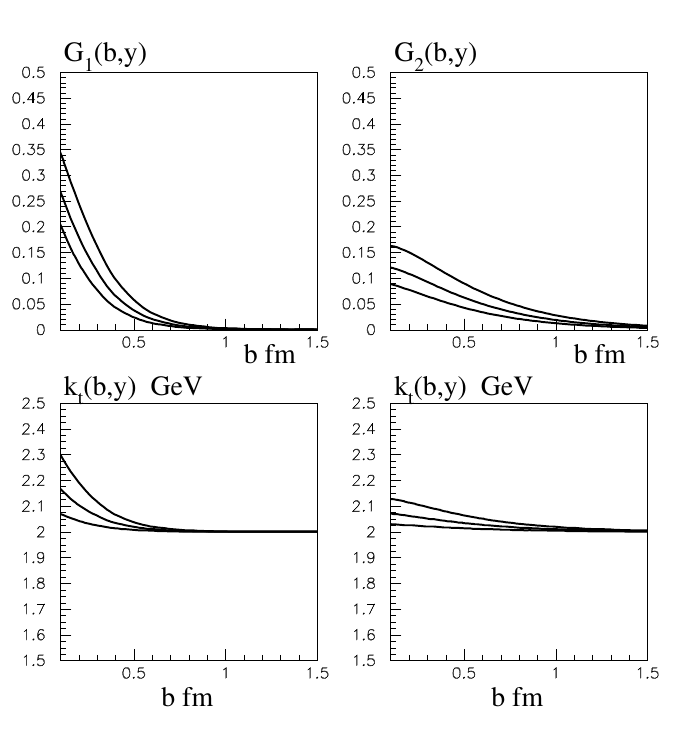}
\caption{\sf Impact parameter, $b$, dependence of the parton densities, $G_i(b,y)$ (upper panels) and the characteristic transverse momenta, $k_{ti}(b,y)$ (lower panels) for the two G-W components, $|\phi_1\rangle$ (left) and $|\phi_2\rangle$ (right) at three values of rapidity $y=$ 9, 6, 3 -- the curves from top to bottom. We use the values of the parameters which have been tuned to describe the total and elastic $p\bar p$ and $pp$ cross sections in the S$p\bar p$S, Tevatron and the LHC colliders energy range. }
\label{f3}
\end{center}
\end{figure}

The $b$ dependence of $G(b,y)$ and $k_t(b,y)$ at few values of $y=3,~6$ and 9 generated by this model is shown in Fig.\ref{f3} where we have used the parameters tuned to describe the total and elastic $p\bar p$ and $pp$ cross sections in the S$p\bar p$S, Tevatron and the LHC colliders energy range, as described in sect.4.   

\section{Formulae for observables}
To calculate the cross section of the high energy proton-proton interaction we have to consider the collision of the two parton cascades generated by the incoming beam and target hadrons. We start with the collisions of  $i$ and $j$ G-W  components. The effective opacity
 $\Omega_{ij} $ is given by
\be\label{OAA}
\Omega_{ij}(b_{ij})=\int d^2 b_1d^2b_2 G_i(b_1,y_1)\frac 1{\sigma_0}G_j(b_2,y_2)\delta^{(2)}(\vec b_{ij}-\vec b_1+\vec b_2)\ ,
\ee
where $b_{ij}$ is the transverse separation (impact factor) between the two colliding protons and $y_1+y_2=Y=\ln s$ is the full rapidity interval between the beam and target hadrons. The dimensionful factor $\sigma_0$ accounts for the cross section of elementary parton-parton interaction. Recall that at the beginning of the evolution the probability, $G(b)$, to find a parton at point $b$  was proportional to $\sigma_0$. Therefore to cancel the extra $\sigma_0$ we are required to have $\sigma_0$ in denominator of (\ref{OAA}).

\subsection{Total and elastic cross sections}
The elastic scattering amplitude reads
\be
\label{eik}
A_{ij}(b)~=~i\left(1-e^{-\Omega_{ij}(b)/2}\right)
\ee
leading to a total cross section
\be
\label{tot}
\sigma_{\rm tot}~=~2\int d^2b\sum_{ij}|a_i|^2|a_j|^2\left(1-e^{-\Omega_{ij}(b)/2}\right) \,
\ee
and a differential elastic cross section
\be
\frac{d\sigma_{\rm el}}{dt}~=~\frac{1}{4\pi}  \left| \int d^2b~e^{i\q_t \cdot \b} \sum_{i,j}|a_i|^2 |a_j|^2~(1-e^{-\Omega_{ij}(b)/2}) \right|^2,
\label{el}
\ee
where $t=-|q_t|^2$. 
The $t$ slope of the elastic cross section $B_{\rm el}$ at $t=0$ can be calculated as the mean $\langle b^2\rangle$. That is
\be
B_{el}(t=0)~=~\frac{\left| \int d^2b~b^2 \sum_{i,j}|a_i|^2 |a_j|^2~(1-e^{-\Omega_{ij}(b)/2}) \right|^2}
{\left| \int d^2b~ \sum_{i,j}|a_i|^2 |a_j|^2~
(1-e^{-\Omega_{ij}(b)/2}) \right|^2}\, .
\label{Bel}
\ee

Formally the result should not depend on the rapidity $y_1$ at which the collision of the two parton cascades was calculated. Our simplified model does not fulfill this condition exactly. However the results do not depend too much on the particular  $y_1$ value. If, for example, instead of the usual $y_1=y_2=Y/2$ we take $y_1=Y/8$ and $y_2=7Y/8$ then the values of $\sigma_{\rm tot}$ change by less than 6\% and the elastic slope $B_{\rm el}$ by less than 1\%.

\begin{figure} [t]
\begin{center}
\includegraphics[trim=-3.0cm 0cm -3cm 0cm,scale=0.6]{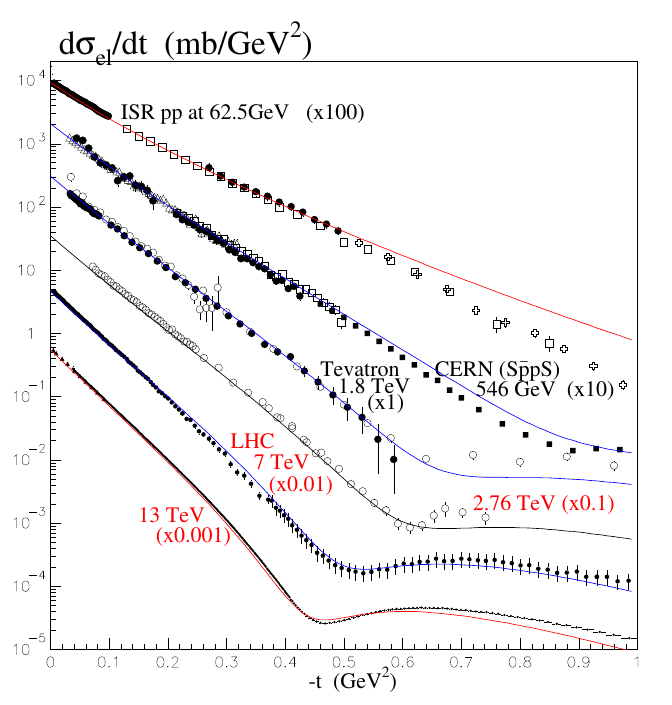}
\caption{\sf The $t$ dependence of the elastic proton-proton (proton-antiproton) cross sections in the S$p\bar p$S, Tevatron and the LHC colliders energy range. The parameters of model were tuned as described in sect.4. The data are taken from \cite{elastic}.  The poor description of the data at the larger values of $-t$ can be improved by using a more detailed G-W parameterization, but this is not relevant to our study.}
\label{f4}
\end{center}
\end{figure}

Up to now we have calculated just the imaginary part of the amplitude. Since we are dealing with the even-signature amplitude~\footnote{The odd-signature contributions are not included in the evolution.} the real part can be restored via dispersion relations. In our high energy limit we use it for fixed $b$ (i.e. for a fixed partial wave with orbital angular momentum $l=b\sqrt s/2$) in the simplified form
\begin{equation}
\label{disp}
\mbox{Re}A(b,s)~=~\frac \pi 2\frac{\partial\mbox{Im}A(b,s)}{\partial\ln s}\, .
\end{equation}
  This real part has been included in the results presented in Fig.\ref{f4}.

\subsection{High-mass diffractive dissociation}
To obtain the cross section of diffractive dissociation we have to consider the case where in the rapidity interval from $y_1$ to $Y$ we have elastic scattering (upper part of the diagram in Fig.\ref{f1}) while below $y_1$ there is an inelastic process (in Fig.\ref{f1} it is shown by the lower central Pomeron). Besides this we have to include the gap survival factor, $\exp(-\Omega_{ij}/2)$ for the amplitude, to be sure that there are no additional inelastic interactions which may fill the gap.

The corresponding cross section takes the form
$$\frac{\xi d\sigma^{\rm SD}}{d\xi}~=~\frac{d\sigma^{\rm SD}}{dy_1}~=
~\int d^2b_1\sum_j |a_j|^2\frac{\lambda G_j(b_1,y_1)}{\sigma_0}d^2b_2 $$
$$\cdot\left(
\sum_i |a_i|^2(1-\sqrt{1-G_i(b_2,y_2)})e^{-\Omega_{ij}(\vec b_1+\vec b_2,Y)/2}S^{\rm enh}_{i}(b_2,y_1)\right) $$
\begin{equation}
\label{dis}
~\cdot
\left(\sum_{i'} |a_{i'}|^2(1-\sqrt{1-G_{i'}(b_2,y_2)})e^{-\Omega_{i'j}(\vec b_1+\vec b_2,Y)/2}S^{\rm enh}_{i'}(b_2,y_1)\right)^*,
\end{equation} 
where $y_2=Y-y_1$ and the `elastic' amplitude $(1-e^{-\Omega/2})$ generated by the parton cascade (in the upper part of Fig.\ref{f1}),   $G_i(b_2,y_2)=1-\exp(-\Omega_i(b_2,y_2))$ is written as $(1-\sqrt{1-G})$.

Recall that $\lambda=g_{3P}/g_N$ is the ratio of the triple-Pomeron to Pomeron-nucleon couplings.
Its value determines the probability of interactions within a unit interval of rapidities. Thus $\lambda$ is proportional to the parton density in rapidity evolution which in its turn is of the order of $\Delta$. 

\begin{figure} [t]
\begin{center}
\includegraphics[trim=-3.0cm 0cm -3cm 0cm,scale=0.7]{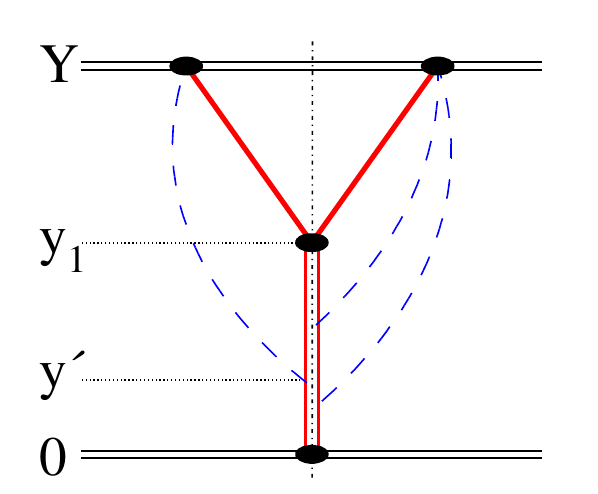}
\caption{\sf Enhanced diagrams (shown by the dashed blue lines) which describe the probability of LRG survival with respect to the interactions with the intermediate partons. }
\label{f:enh}
\end{center}
\end{figure}

Finally the factor $S^{\rm enh}_{ij}(b_2,y_1)$ accounts for the probability of LRG survival with
 respect to soft interactions with the intermediate partons from the cut Pomeron~\footnote{`Cut' denotes the Pomeron which produces the secondary hadrons (like that in the lower part of Fig.\ref{f1}a) and not the Pomerons which describe the absorptive effects (like that shown by dashed curves in Fig.\ref{f2}).} (in the lower part of Fig.\ref{f1}). It is given by the sum of the enhanced diagrams (see the dashed blue lines in Fig.\ref{f:enh})
\begin{equation}
\label{Senh}
S^{\rm enh}_{i}(b,y_1)=\exp\left(-\int_{1.6}^{y_1}dy'\frac\lambda 2 G_i(b,Y-y')\right)\ .
\end{equation}
Here we start the integration over $y'$ from $y'_{\rm min}=1.6$ since 
 the interval of lower $y'$ is already accounted for in terms of the G-W eigenstates.

Strictly speaking there should be the integration over the position of the new interaction point $b$ in the impact parameter plane. However, since due to the large value of $k_0\sim 2$ GeV (i.e. the  small slope of the Pomeron trajectory $\alpha'_P$) the diffusion in the $b$ plane is rather weak, we put in(\ref{dis}) a fixed value of $b=b_2$~\footnote{This, second order, effect of the diffusion in the $\vec b$ plane should be accounted for in a future more precise version of the model.}.

Next, 
the slope of diffractive dissociation $B_{\rm dis}(t=0)=\langle b^2_2 \rangle$ reads
\begin{equation} 
\label{Bdis}
B_{\rm dis}(t=0)~=~\frac{\int d^2b_1\sum_j |a_j|^2
G_j(b_1,y_1)d^2b_2b^2_2\left(\sum_i |a_i|^2 ...
\right)\left(\sum_{i'}|a_{i'}|^2...\right)^*}{\int d^2b_1\sum_j |a_j|^2 G_j(b_1,y_1)d^2b_2\left(\sum_i |a_i|^2 ...
\right)\left(\sum_{i'}|a_{i'}|^2...\right)^*}\ ,
\end{equation}
where `dots' denote the corresponding expressions in the second and third lines of (\ref{dis}).

\subsection{Density of secondaries in LRG events}
\begin{figure} [t]
\begin{center}
\includegraphics[trim=-3.0cm 0cm -3cm 0cm,scale=0.7]{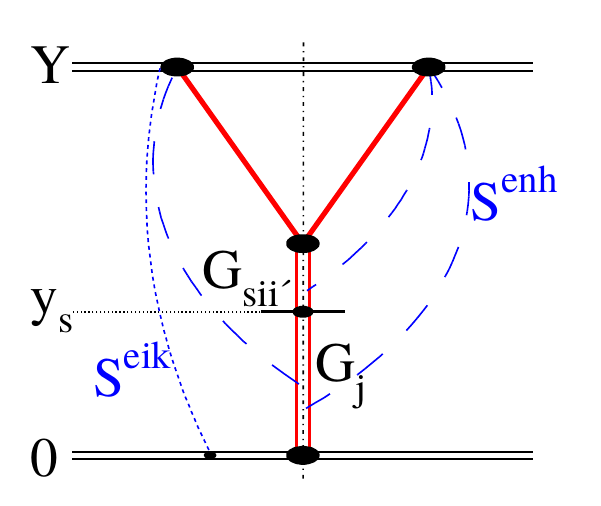}
\caption{\sf The diagram for the inclusive one  particle cross section for SD events. Screening effects are indicated by the (blue) short-dashed ($S^{\rm eik})$ and (blue) long-dashed ($S^{\rm enh}$) curves which describe the probability of LRG survival with respect to additional proton-proton interactions or the interactions with the intermediate partons. }
\label{yenh}
\end{center}
\end{figure}

The inclusive cross section of secondaries produced at rapidity $y_s$ in the high-mass dissociation is (see Fig.\ref{yenh})

\begin{equation}
\label{dis-y1}
\frac{\xi d\sigma^{\rm SD}}{d\xi dy_s}=\int d^2b_1 d^2b_s\sum_{iji'}|a^2_i| |a_{i'}^2||a_j|^2
\frac{g_sG_j(b_1,y_s)G_{sii'}(b_s,y_3)}{\sigma_0}S^{\rm eik}_{ij}S^{\rm eik}_{i'j}S^{\rm enh}_{ii'}(b_s,Y-y_{\rm gap})\ ,
\end{equation}
where the constant $g_s$ is the probability of secondary particle emission from a one cut Pomeron. We put $g_s=2.2$ in order to have the density of charged particles in non-diffractive events $dN_{\rm ch}/dy=6$ at $\sqrt s=13$ TeV to be in agreement with the data.
Note that here we introduce an additional Green's function, $G_s(b_s,y_3)$, which describes the development of the parton cascade within the rapidity interval $y_3$ between the triple Pomeron vertex (at $y_{\rm gap}=-\ln\xi=Y-y_1$) and the new produced particle (placed at $y_s$ and $b_s$ in rapidity and impact parameter plane); so $y_3=y_{\rm gap}-y_s$. This function satisfies the same evolution equations (\ref{emast}) and (\ref{ekt}) as $G_j$
but with the initial conditions 
\begin{equation}
\label{dis-y2}
G_{sii'}(b_s,0)=\lambda\left(1-\sqrt{1-G_i(b_s,y_{\rm gap}})\right)\left(1-\sqrt{1-G_{i'}(b_s,y_{\rm gap}})\right)
\end{equation} 
and
\begin{equation}
\label{kt-dis}
k_{t;sii'}(b_s,0)~=~\sqrt{k_{t,i}(b_s,y_{\rm gap})k_{t,i'}(b_s,y_{\rm gap})}\ ,
\end{equation}
where $k_{t,i}$ and $k_{t,i'}$ are the values of $k_t$ of the G-W components 
$i$ and $i'$ respectively.

The gap survival factors $S^{\rm eik}$ account for the incoming proton interactions
\begin{equation}
S^{\rm eik}_{ij}=e^{-\Omega_{ij}(\vec b_1+\vec b_s,Y)/2}
\end{equation}
while the value of $S^{\rm enh}_{ii'}$ is given in terms of (\ref{Senh}) as
\begin{equation}
S^{\rm enh}_{ii'}=S^{\rm enh}_i(b_s,Y-y_{\rm gap})S^{\rm enh}_{i'}(b_s,Y-y_{\rm gap})\ .
\end{equation}
 The corresponding opacity
 $\Omega_{si} $ can be calculated via
 $$\exp(-\Omega_{si}(b_{s},Y-y'))=1-G_i(b_s,Y-y')\ .$$

\subsection{Parton transverse momenta}
In order to evaluate the characteristic transverse momenta of the  secondaries produced at some rapidity $y'$ we can  multiply by $k_t(b,y')$ (\ref{ekt}) the value of $G(b,y)$   for each G-W component $j$ at $y=y'$ and then continue the evolution of this product $G_{(k_t),j}(b,y)=k_t(b,y')_jG_j(b,y)$
according to the master equation (\ref{emast}). 
The mean value $\langle k_t(y')\rangle$ is given by the ratio of `cross sections' (say, (\ref{dis})) calculated with $G_{(k_t),j}(b,y)$ to that calculated with the normal $G_j(b,y)$. Of course this $\langle k_t(y') \rangle $ is not equal to the mean momentum of the secondary hadrons, $\langle p_t\rangle$,  which can be measured experimentally. The value of $\langle p_t\rangle$ will be modified by hadronization. However by looking at the energy, rapidity and $b$ dependences of $\langle k_t\rangle$ we get some semi-quantitative understanding of the  expected
 $\langle p_t\rangle$ behaviour.

 \subsection{Secondary Reggeon contributions}
Besides the triple-Pomeron (PPP) term  considered in sect.3.2 there are the contributions caused by secondary Reggeon (R) exchange. For  relatively large $\xi$ (that is not too large $y_2$) one has to account for the RRP term where the two upper Pomerons in Fig.\ref{f1}b are replaced by R-exchange.

Assuming that the R-reggeons are emitted
from valence quarks for all G-W eigenstates we put the {\em same} vertex couplings and form factors and write the corresponding exchange amplitude as
\begin{equation}
\label{Ra}
A_R(b)=1-e^{\Omega_R(b,y_2)/2}
\end{equation}
with
\begin{equation}
\label{Ro}
\Omega_R(b,y_2)~=~\frac{\sigma_R~ e^{(\alpha_R(0)-1)y_2}
}{4\pi B_R}~e^{-b^2/4B_R}\, .
\end{equation}
We take the intercept of the R-trajectory to be $\alpha_R(0)=1/2$ and the slope $B_R=\alpha'_R y_2+2/0.71$ GeV$^2$ with $\alpha'_R=0.9$ GeV$^{-2}$; the term $2/0.71$ corresponds to the dipole form factor $F_R(t)=1/(1-t/0.71)^2$. 

Thus for the RRP contribution we obtain
\begin{equation}
\label{disR}
\frac{\xi d\sigma^{RRP}}{d\xi}=\int d^2b_1\sum_j |a_j|^2
\frac{G_j(b_1,y_1)}{\sigma_0}d^2b_2\left|\sum_i |a_i|^2(1-e^{-\Omega_R(b_2,y_2)/2})S^{\rm eik}_{ij}
S^{\rm enh}_i(b_2,Y-y_2)\right|^2,
\end{equation} 

For very small $\xi$ corresponding to low mass, $M_X$, dissociation the central Pomeron (in the lower part of Fig.\ref{f1}) can be replaced by a R-reggeon. This forms the PPR term whose contribution decreases as $1/M_X\propto \exp(-y_1/2)$. However this, relatively low $M_X$, contribution in our case was accounted for within the G-W formalism. To obtain a more or less realistic behaviour at the lowest $\xi$ end we assume resonance - `reggeon exchange'  duality and 
redistribute the low mass dissociation cross section given by (\ref{eq:b3}) (minus the elastic cross section (\ref{eq:b2})) over $y_1$ with a $0.5\exp(-y_1/2)$ weight.

In each case the corresponding $t$-slope was calculated as the mean value $\langle b^2_2\rangle$.

\subsection{Multiplicity distribution }
The multiplicity distribution of charged hadrons observed in some rapidity interval is given by the convolution of several  functions. First, this is the distribution of secondaries produced by one individual `cut' Pomeron. It includes 
the distribution over the number of $s$-channel gluons and the effects of hadronization. Next we have the distribution over the number of Pomerons. Finally, the result may be affected by the "colour reconnection" between the gluons from different Pomerons. 

In the present model we neglect the colour reconnection effects and assume that the charged particles are emitted by one Pomeron according to Poisson's law. To account for the charge conservation we take the Poisson over the number, $N_1=N_1^+$, of positively charged particles. The element which will be studied below is the effect on the multiplicity distribution coming from the number, $n$, of the Pomerons. 

In the one-channel eikonal approximation, that is for each G-W component,  the distribution over the number of Pomerons  also takes a Poisson form
\begin{equation}
\label{PP}
P_P(n)~=~\frac{\Omega^n(b)}{n!}e^{-\Omega(b)}\, ,
\end{equation}
 where the mean number of the  cut Pomerons, $\langle n\rangle=\Omega(b)$, depends on particular $b$ value. That is actually we deal with the sum (integral) of a continuous number of Poissons with different $\langle n(b)\rangle$. This leads to the final distribution
\begin{equation}
\label{Ph}
P_h(N)~=~\int  w(b)\sum_n\frac{\Omega^n(b)}{n!}~e^{-\Omega(b)}~\frac{(n\cdot N_1)^N}{N!}~e^{-n\cdot N_1}\ d^2b\, ,
\end{equation}
where the weight $w(b)$ is given by the integrand of the corresponding cross section. For non-diffractive inclusive events 
$$w_{ij}(b)~=~\frac{1-\exp(-\Omega_{ij}(b))}{\int d^2b\ (1-\exp(-\Omega_{ij}(b))}$$
while for high-mass diffractive dissociation (\ref{dis}) 
\begin{equation}
\label{w-dis}
w_{j}=\frac{G_j(b_1,y_1)\ \int d^2b_2~|...|^2}{\int d^2b_1 G_j(b_1,y_1)\int d^2b_2~|...|^2}\, ,
\end{equation}
where for simplicity we consider just a collision of a particular ($i$ and $j$) G-W eigenstates; $|...|^2$ denotes the last two factors on the r.h.s. of (\ref{dis}).  The $\Omega(b)$ which should be used in (\ref{Ph}) is equal to
$\Omega_j(b_1,y_1)=-\ln(1-G_j(b_1,y_1))$. 

\section{Parameters of the model}
Let us, first, discuss the expected reasonable values of the parameters of our model.

The free parameters which are used to tune the model 
are:
\begin{itemize}
\item The Regge intercept of the original (unscreened) Pomeron, $1+\Delta$; from NLL BFKL we expect $\Delta\sim 0.2$.
\item
The initial value of the parton transverse momentum, $k_0=k_t(b,y=0)$.\\
$1/k^2_0$ plays the role of the slope, $\alpha'_P$, of the Pomeron trajectory. This slope is known to be rather small, 
say, $\alpha'_P=0.25$ GeV$^{-2}$ in the  parametrization of ~\cite{DL}. Even a smaller  $\alpha'_P=0.14$ GeV$^{-2}$ was obtained in~\cite{Burq}. Thus $k_0\sim 2$ GeV looks to be reasonable value.
\item
Next, we have the elementary wee parton cross section, $\sigma_0$, which should be of the order $2\pi/k^2_0\sim 1$ mb.
\item Finally, we have the initial impact parameter distribution of the partons in each G-W eigenstate, which in our simplified model are described by a total of 6 parameters, as explained below.
\end{itemize}

Since we are looking mainly for the qualitative and semi-quantitative effects we try to be as simple as possible and take only two G-W components with equal weight $a_1=a_2=1/\sqrt 2$. For each of these two G-W eigenstates the $b$ dependence is parameterised by factors of the form
\be
F_i(t)={\rm exp}(-\sqrt{d_i(c_i-t)}+\sqrt{d_ic_i}),
\label{eq:ff}
\ee
where $c_i$ is added to avoid a singularity at $t=0$. Note that $F_i(0)=1$.   The starting distributions for the evolution in rapidity are
\begin{equation}
G_i(b,y=0)~=~\frac{f_i}{4\pi}\int dt J_0(b\sqrt |t|)F_i(t)\ .
\end{equation}
Thus we have 3 free parameters ($f_i$ which determines the value of the  parton density, $d_i$ and $c_i$)  for each G-W eigenstate.

The values of parameters found to describe the data are listed in Table 1.
  \begin{table}[htb]
\begin{center}
\begin{tabular}{|l|c|}\hline
  $\Delta$  & 0.17 \\
  $\sigma_0 $ (GeV$^{-2}$)  &1.18\\
    $k_0~(\GeV)$   & 2.2 \\
    $\lambda=g_{3P}/g_N$ & 0.2 (fixed)\\
   \hline
  $f_1$ &  11\\
  $ d_1~(\GeV^{-2})$   & 2.75 \\
  $ c_1~(\GeV^2) $ & 0.2 \\
  $f_2$ & 4.15\\
  $ d_2~(\GeV^{-2})$   & 1.3 \\
  $ c_2~(\GeV^2) $ & 0.3 \\
 \hline

\end{tabular}
\end{center}
\caption{\sf The values of the parameters in the two-channel eikonal fit to elastic $pp$ ($p\bar p$) scattering data.
}
\label{tab:1}
\end{table}
The first two parameters in Table 1 control the absolute value and the energy behaviour of the total cross section.
$k_0$ is responsible for the shrinkage of diffractive cone, that is - for velocity of diffusion in $b$ space while $\lambda$ determines the probability of high mass diffractive dissociation. We fix $\lambda=0.2$ to be equal to the value given by both - the analysis based on the perturbative QCD approach and the HERA data~\cite{Em} and the triple-Regge analysis accounting for absorptive corrections~\cite{Luna}. 
The final 6 parameters define the parton densities and their $b$ distribution in the two G-W eigenstates.

The parameters were tuned to reasonably describe the elastic $pp$ ($p\bar p$) cross sections in the collider energy range as shown in Fig.\ref{f4}. As a rule, when tuning the parameters, we use only two digits~\footnote{Thus it may be possible to improve the description.}, since our goal is not to obtain the most precise description, but instead to achieve a qualitative understanding of the multi-Pomeron contributions and a semi-quantitative evaluation of the expected effects. In other words, we are seeking a general understanding of how high energy diffractive phenomena are driven by perturbative QCD.  The fact that the values found for the parameters turn out to be in agreement with preliminary qualitative expectations gives support for the model.

The resulting cross sections and elastic slope are presented in Table 2. Note that the model gives a reasonable probability of low-mass diffractive dissociation, $\sigma^{\rm SD}_{\rm lowM}=3.75$ mb at $\sqrt s=7$ TeV in agreement with the TOTEM, $\sigma^{\rm SD}_{\rm lowM}=2.6\pm 2.2$ mb,~\cite{TOT-lM} measurement.

\begin{table}[htb]
\begin{center}
\begin{tabular}{|c|c|c|c|c|c|}\hline
 $\sqrt{s} $ &    $\sigma_{\rm tot} $ & $\sigma_{\rm el} $ &  $B_{\rm el}(t=0)$  &   $\sigma^{\rm SD}_{\rm lowM} $\\
 \hline
 
  (TeV)   &           mb    &     mb    &     (GeV$^{-2}$)  &  mb\\
\hline
    0.0625  &    42.5   &     7.3   &    11.6   & 1.46\\
    0.546     &   63.9   &    13.5   &    14.7   &  2.40  \\
   1.8  &      78.1   &    18.0   &    16.8  &   3.03\\
   7  &      96.4   &    24.2   &    19.6    &    3.82\\
   8   &      98.3   &    24.8    &   19.9    &    3.90\\
  13   &    105.5    &   27.3    &   21.0     &   4.21\\
\hline

\end{tabular}
\end{center}
\caption{\sf The predictions of the elastic and diffractive observables resulting from the description of the presently available data. }
\label{tab:2}
\end{table}

\section{Results for diffractive dissociation} 
 \subsection{Cross section of single proton dissociation}
\begin{figure} [t]
\begin{center}
\includegraphics[trim=-3.0cm 0cm -3cm 0cm,scale=0.6]{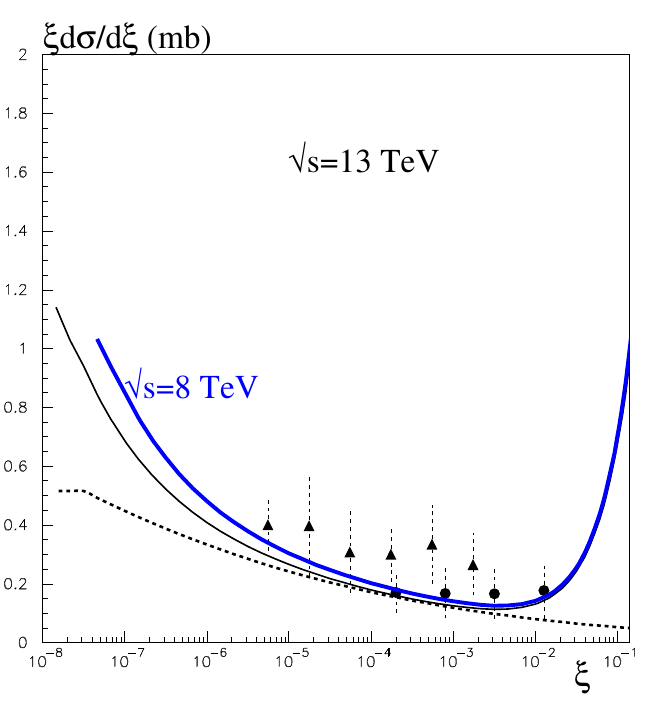}
\caption{\sf The $\xi$ dependence of the single dissociation (SD) cross section at $\sqrt s=13$ TeV (black). The dashed curve is the  Pomeron component while the continuous curve includes the secondary Reggeon contribution. ATLAS (8 TeV)~\cite{ATL} and CMS (7 TeV)~\cite{cms} data are shown by circles and triangles respectively. The CMS points have been reduced by a factor of 1.27 to  approximately account for the fact that 
 these data contain some admixture of double dissociation, in addition to pure SD~\cite{NM127}. Thick blue/upper curve corresponds to $\sqrt s=8$ TeV. At $\xi>3\cdot 10^{-5}$ it is very close to  the black curve.}
\label{f5}
\end{center}
\end{figure}

 The expected $\xi$ behaviour of the cross section of single proton dissociation (SD) is shown in Fig.\ref{f5}. The pure Pomeron component is shown by the dashed curve while the solid curve includes the secondary reggeon contribution (as described in sect.3.5). The black curves correspond to $\sqrt s=13$ TeV. The result for $\sqrt s=8$ TeV is shown by the thick blue curve. Here we use $\lambda=0.2$~\cite{Em,Luna} and $\sigma_R=22$ mb which is consistent with the analysis of~\cite{Luna} and the secondary Reggeon contribution in the COMPETE fit~\cite{COMPETE} of the $pp$ total cross sections. 
 
\begin{figure} [t]
\begin{center}
\includegraphics[trim=-3.0cm 0cm -3cm 0cm,scale=0.6]{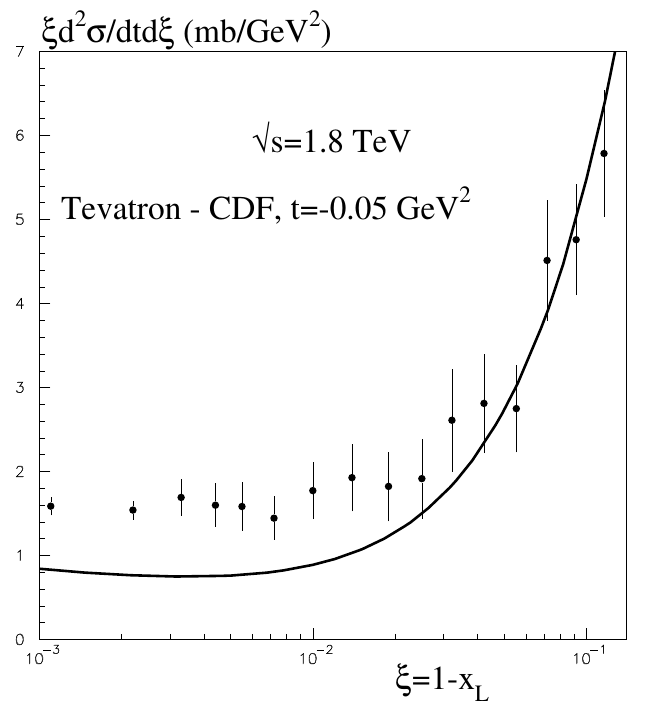}
\caption{\sf The comparison of the model with the 
results of the analyses by Goulianos 
and Montanha
of the CDF data ~\cite{cdf} at $\sqrt s=1.8$ TeV and $t=-0.05$ GeV$^2$.}
\label{f6}
\end{center}
\end{figure}

 Recall that there is some tension between the points extracted by Goulianos and Montanha~\cite{cdf} from the CDF data  and the cross sections of diffractive dissociation measured at the LHC. With $\lambda=0.2$ we underestimate the CDF cross section at $\xi<0.01$ (see Fig.\ref{f6}) 
but overshoot a little the recent  ATLAS~\cite{ATL} results (see~\cite{KMR14} for a discussion).

\begin{figure} [t]
\begin{center}
\includegraphics[trim=-3.0cm 0cm -3cm 0cm,scale=0.6]{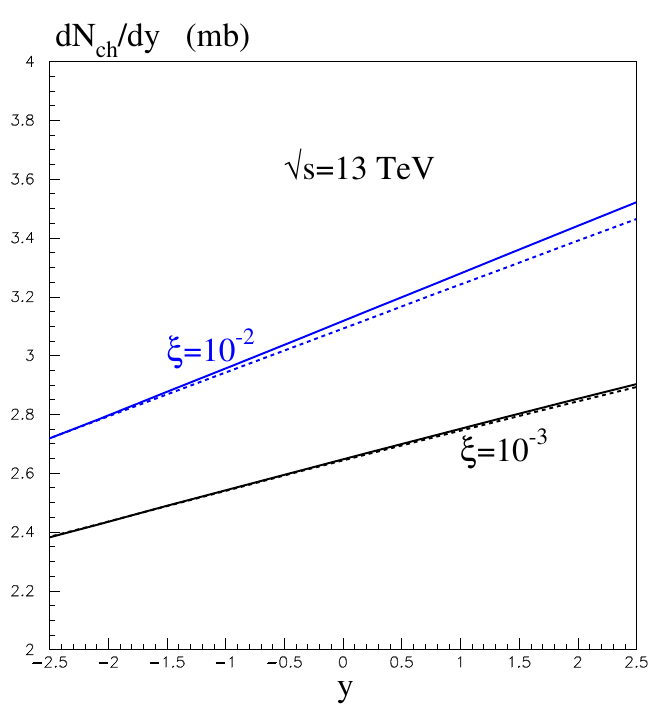}
\caption{\sf The rapidity dependence of the charged multiplicity observed in the central detector for SD events with $\xi=0.01$ (blue) and 0.001 (black) at $\sqrt s=13$ TeV.  The dashed curves correspond to the pure Pomeron-induced cross section without the secondary Reggeon contribution. }
\label{fy}
\end{center}
\end{figure}

\begin{figure} [t]
\begin{center}
\includegraphics[trim=-3.0cm 0cm -3cm 0cm,scale=0.6]{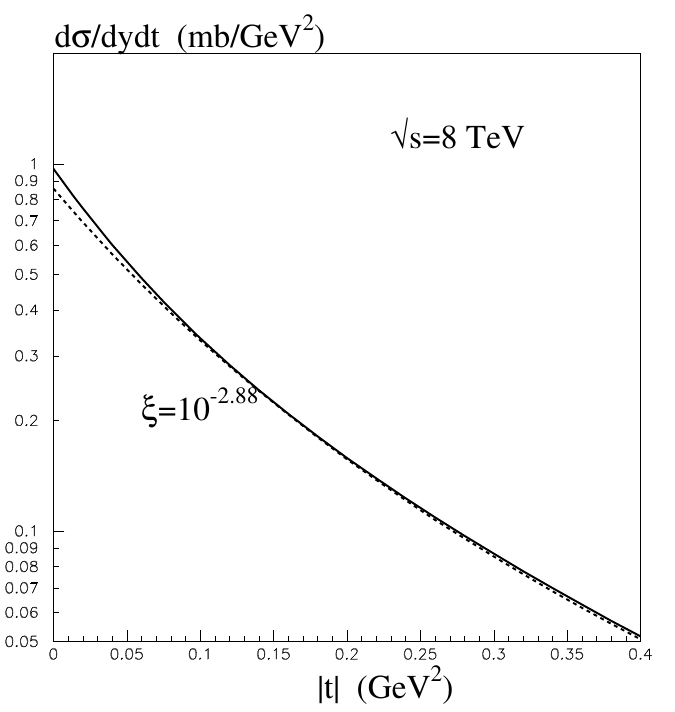}
\caption{\sf The $t$ dependence of the SD cross section $d\sigma/dtdy$ at $\sqrt s=8$ TeV and $\xi=10^{-2.88}$ (this value of $\xi$ is chosen to compare with the ATLAS-ALFA~\cite{ATL} results at $\langle \xi\rangle=10^{-2.88}$). The dashed curve corresponds to the pure Pomeron-induced cross section without the secondary Reggeon contribution. }
\label{ft}
\end{center}
\end{figure}

For $\xi < 0.01$ (where the RRP contribution becomes small) the value of $d\sigma^{\rm SD}/d\ln\xi$ increases with decreasing $\xi$  mainly due to the Pomeron intercept $1+\Delta > 1$. However this growth is tamed by absorptive effects. At very small $\xi$, corresponding to low $M_X$, we see the contribution of the PPR term coming from G-W low-mass dissociation.

\subsection{Rapidity distributions} 
We show in Fig.\ref{fy} the rapidity dependence of the charged particle densities $dN_{\rm ch}/dy$  expected in SD events in the central detector interval. Contrary to the standard plateau observed in this region in the non diffactive events the particle density $dN_{\rm ch}/dy$ in SD decreases when the rapidity of the secondary meson approaches the edge of the LRG (i.e. to the position of the triple-Pomeron vertex). This behaviour can be explained by looking at the product $G_j(b_1,y_s)G_{sii'}(b_s,y_3=y_{\rm gap}-y_s)$ in (\ref{dis-y1}). Indeed, near the gap edge we deal with the beginning of the $G_{sii'}$ evolution where the particle density is rather small and the value $G_{sii'}$ increases rapidly. On the other hand the function $G_j(b_1,y_s)$ is already close to saturation  and weakly depends on $y_s$ (here $y_s$ is large). Therefore the product $G_j(b_1,y_s)G_{sii'}(b_s,y_3=y_{\rm gap}-y_s)$ increases with $y_3$, i.e. decreases when $y_s$ approaches the gap edge $y_{\rm gap}$.

 \subsection{$t$ dependence of SD cross section}
 The $t$-dependence of the SD amplitude can be calculated via the Fourier
transform over the impact parameter $b_2$ (in (\ref{dis})). Except for very small $|t|$ the distribution is rather close to a simple
exponent (see Fig.\ref{ft} as an example).
\begin{figure} [t]
\begin{center}
\includegraphics[trim=-3.0cm 0cm -3cm 0cm,scale=0.6]{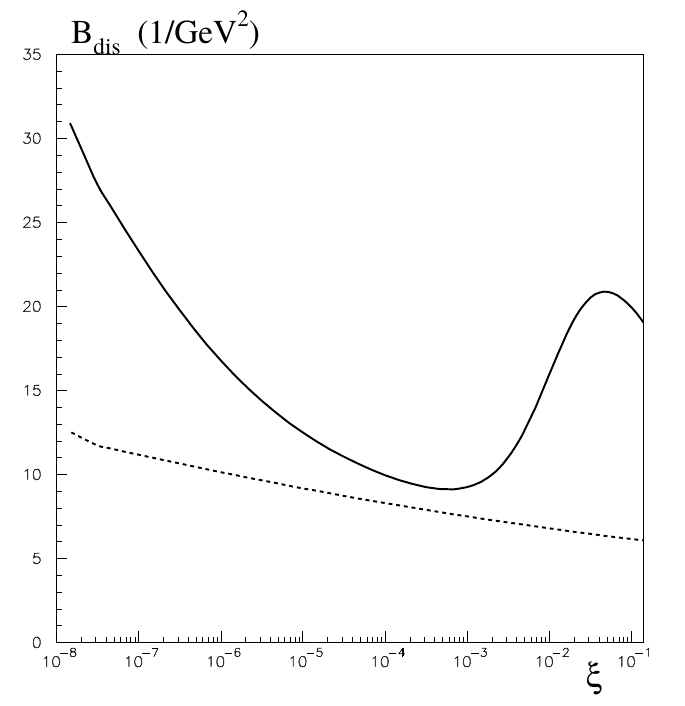}
\caption{\sf The $\xi$ dependence of the $t$-slope $B_{\rm dis}(t=0)$ in the single proton
dissociation process at $\sqrt s=13$ TeV. The dashed curve is the  Pomeron
component while the continuous curve includes secondary Reggeon
contributions. Note that here we show the slope at $t=0$. As it is seen
from Fig.\ref{ft} the mean slope (within a larger $|t|$ interval) is a bit
smaller.}
\label{f7}
\end{center}
\end{figure}

\begin{figure} [t]
\begin{center}
\includegraphics[trim=-3.0cm 0cm -3cm 0cm,scale=0.6]{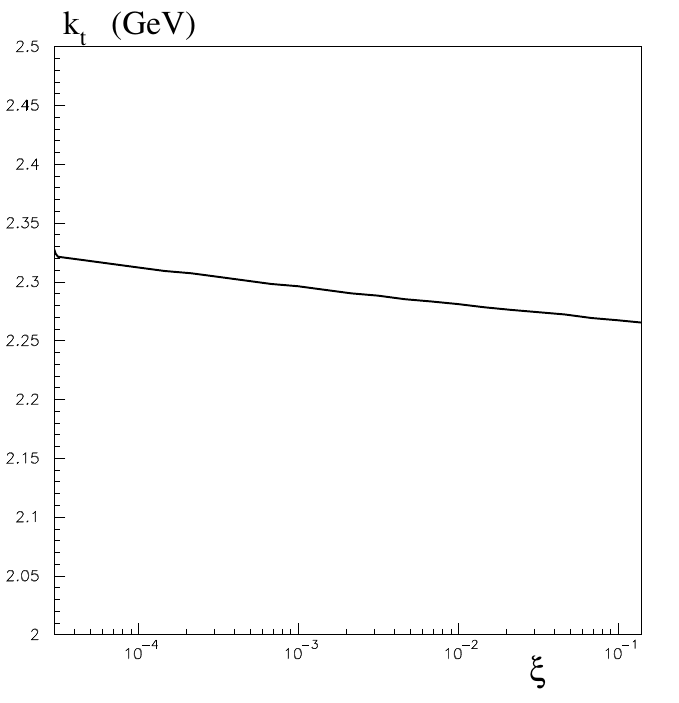}
\caption{\sf The $\xi$ behaviour of the characteristic transverse momentum
$k_t$ measured at $\eta=0$ (in the laboratory frame, i.e. the $pp$ centre of mass, system)  in single proton dissociation at
$\sqrt s=13$ TeV.}
\label{f8}
\end{center}
\end{figure}

The value of the slope expected in proton diffractive dissociation is
shown in Fig.\ref{f7}. Note that the secondary Reggeon terms enter with a
very large slope $B_{\rm dis}(t=0)$ (up to 40 GeV$^{-2}$ at $\xi=10^{-3}$).
Therefore for $\xi\gapproxeq 0.003$ (where the role of the secondary RRP
contribution becomes  important) the value of $B_{\rm dis}$ increases with
$\xi$.   The large value of $B_{\rm dis}$ in RRP term is explained by strong
absorption which pushes the PPR and RRP contributions to the far periphery
of the disk. So, only the large $b_2$ tail  survives.

On the other hand the slope corresponding to the pure Pomeron-induced
dissociation is  smaller ($B_{\rm dis}\simeq 7$ GeV$^{-2}$ at $\xi=10^{-3}$).
In this case the large $b_t$ needed to go to the periphery of the disk is
mainly provided by a large
  $b_1$ corresponding to the central (in Fig.\ref{f1}) ``inelastic" (cut)
Pomeron while the value of $b_2$ (responsible for the  interaction 
the LRG) stays rather small. If we neglect the enhanced diagrams in
Fig.\ref{f:enh} then we get $B_{\rm dis}\simeq 5$ GeV$^{-2}$ (at $\xi=10^{-4} -
10^{-3}$). Only the $S^{\rm enh}$ survival factor allows a larger
$B_{\rm dis}$ up to 7 - 8 GeV$^{-2}$ by
  absorbing part of the low-$b_2$ contribution.
  
  The growth of $B_{\rm dis}$ at very small $\xi$ is due to the slope of the effective
Pomeron trajectory, $\alpha'_{P, \rm eff}$, (i.e. expansion of the
disk in $b$ space) and the PPR term which describes low-mass
dissociation.
  We emphasize  that in Fig.\ref{f7} we have plotted the slope at $t=0$,
which is larger than the mean slope $B_{\rm dis}$ fitted in some finite
$t$-interval. In particular at $\sqrt{s}$= 8 TeV and $\xi=10^{-2.88}$
the mean slope `measured' between $t=$ 0.02 and 0.32 GeV$^2$ is
$B_{\rm mean}(0.02-0.32)=7.6$ GeV$^{-2}$ while the value of
$B_{\rm dis}(t=0)=10.3$ GeV$^{-2}$.

Formally we have the possibility to introduce some additional slope
$B_{3P}$ of the triple-Pomeron vertex. However its natural value should
be $B_{3P}\sim 1/k^2_0\sim 0.25$ GeV$^{-2}$ which is rather small. On the other
hand the value of $k_0$ controls the shrinkage of the diffractive cone and it
is needed to keep $k_0\sim 2$ GeV in order to reproduce the available
elastic $d\sigma_{el}/dt$ data.

Finally, we show in Fig.\ref{f8} the typical wee-parton transverse momentum at
$\eta_{Lab}=0$ (that is near the centre of mass of the
two colliding protons).  Note that $k_t(\xi)|_{\eta=0}$ weakly
increases with decreasing $\xi$, but still remains close to its initial
value $k_0=2.2$ GeV.
This means that in the diffracted system $X$ we expect the transverse
momentum distribution of secondaries and the mean value of $\langle p_t\rangle$ to be
close to that observed at comparatively low (say, $\sqrt s\sim 20 - 40$
GeV) energies.
The explanation is evident. The dissociation comes mainly from the
periphery of the disk where the parton density is small. Thus, far from
the saturation limit there are no reason to noticeably enlarge $k_t$.
Recall that in non-diffractive inclusive events we get at $\sqrt
s=13$ TeV a  larger $k_t(\eta=0)=2.73$ GeV.

 Note that here we consider the secondaries produced somewhere in the
centre of $M_X$ system and not too close to the edge of LRG. Near the
edge of LRG the situation is more interesting and  complicated.
Recall that  the Pomeron has a small transverse size
(see e.g.~\cite{Pom,Ryskin:2011qh}). In comparison with the proton radius
$\sim 1$ fm the Pomeron size  is $\sim 1/k_0\sim 0.1$ fm. This is
indicated by the small value of the slope of the Pomeron trajectory 
$\alpha'_P\leq 0.25$ GeV$^{-2}$ (see
e.g.~\cite{DL,KMR-slope,GLM})~\footnote{It was shown long ago in terms of
the multiperipheral models~\cite{FC} and in terms of the parton
cascade~\cite{Gr} that the value of $\alpha'\propto 1/k^2_t$ where $k_t$
is the typical transverse momentum of the partons (t-channel propagators
in the case of multiperipheral models). Simultaneously this value of $k_t$
determines the size of the bound system which forms the Regge pole
(Pomeron).}   and the very small (consistent with zero) $t$-slope of the
triple-Pomeron vertex (see e.g.~\cite{KKPT,FF,Luna})~\footnote{In these 
papers the triple Pomeron vertex was extracted fittinng rather old
CERN-ISR data (Tevatron data was included in~\cite{Luna}). However this
energy was sufficient to determine the triple-Pomeron contribution
and since the vertex occupies a limited rapidity interval we can use
the obtained results at  larger  energies, in particular for the LHC
region.}  . Therefore, in comparison with the proton fragmentation region
in `Pomeron fragmentaion' (i.e. near the edge of the LRG) we expect a
larger mean transverse momenta, $p_t$, and a broader $p_t$ distribution of
the secondaries. Some indication in favour of this can be seen in Fig.2
of~\cite{Luk} where  in comparison with the PYTHIA 8 Monte Carlo
simulations the particle density increases with $p_t$.  The `data/MC'
ratio exceeds 1 and reaches about 2 for $p_t>1$ GeV.

Recall also that the Pomeron consists mainly of gluons  and so the Pomeron
is essentially a singlet with respect to the flavour SU(3) group.
Therefore, it would be interesting to observe in the Pomeron fragmentation
region (close to the edge of the LRG) the presence of $\eta$ and $\eta'$
mesons. Since $\eta'$ is almost a singlet of flavour SU(3)
and contains a large gluon component we may expect that the Pomeron
fragmentation region will be enriched by $\eta'$ mesons. Besides this,
there should be a good chance to observe $0^{++}$ and $2^{++}$ glueballs
in the Pomeron fragmentation region.

\subsection{Multiplicity distribution}
As explained in sect.3.6 the expected multiplicity distribution is represented by the sum (integral) of 'Poissons' with different mean $\langle N_{\rm ch}\rangle$ which depends on the particular impact parameter and the number of cut Pomerons.
Since diffractive dissociation events survive only in the region where the probability of multiple parton interactions is small (and correspondingly the multi-Pomeron contributions are suppressed) we expect  in these events a smaller multiplicity and a rather narrow distribution. 
\begin{figure} [t]
\begin{center}
\includegraphics[trim=-3.0cm 0cm -3cm 0cm,scale=0.6]{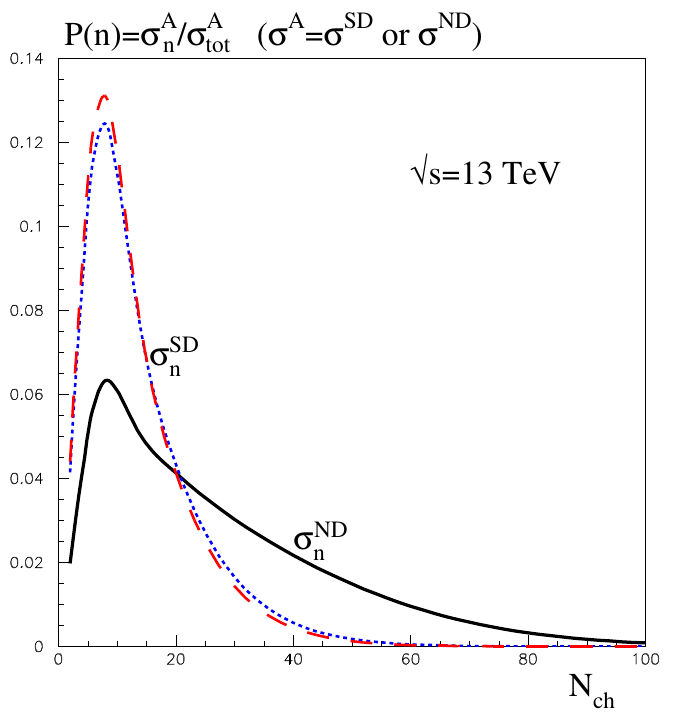}
\caption{\sf The distribution over the charged hadron multiplicity in non-diffractive (ND) events (continuous curve) and in the case of single proton (SD) dissociation  at $\sqrt s=13$ TeV for $\xi=10^{-3}$ (red long-dashed curve) and $\xi=10^{-2}$ (blue short-dashed curve). (We assume that mean number of charged hadrons emitted by {\em one} cut Pomeron is equal to 8 in the rapidity interval that the value of $N_{\rm ch}$ was measured.)}
\label{f9}
\end{center}
\end{figure}
Indeed, as seen in Fig.\ref{f9}, in non-diffractive events we observe a long high $N_{\rm ch}$ tail caused by the integration over a large interval of impact parameters $b$; at each value of $b$ we deal with a different number of cut Pomerons  $\langle n\rangle =\Omega(b)$. On the other hand
 the dissociation events come from the edge of disk where $\Omega(b)\lapproxeq 1$. Therefore here the distribution is  much narrower and the mean multiplicity is about twice smaller ($\langle N^{\rm dis}_{\rm ch}\rangle =13.3$  for $\xi=10^{-3}$ and $\sqrt s=13$ TeV instead of $\langle N^{\rm dis}_{\rm ch}\rangle =27.5$ for the non-diffractive case). 
 
\section{Discussion}
In the previous sections we have studied single diffractive (SD) processes and shown qualitative (semi-quantitative) effects caused by the fact that at small $b$ stronger absorptive corrections push the amplitude of dissociation to the periphery of the disk. In comparison with \cite{KMR11}, where the diffusion in $b$ space was neglected and $\alpha'_P=0$ was assumed, here we pay the most attention just to the possibility that partons move in $b$ plane. On the other hand, in \cite{KMR11} the diffusion in $\ln{k_t}$ was accounted for more precisely. In the present model we consider just the evolution (with rapidity) at a typical value of $k_t$. However, since on the periphery of disk, from which the major SD contribution comes, the parton density is relatively small and the value of $k_t$ practically does not change  we believe that the present model is more appropriate for  analysis the SD processes.

\section*{Acknowledgments}
 We thank Kenneth Osterberg, Paul Newmann, Valery Schegelsky and Marek Tasevsky for discussions.
 MGR thanks the IPPP at the University of Durham for hospitality.

E.~A.~Kuraev, L.~N.~Lipatov and V.~S.~Fadin,
  Sov.\ Phys.\ JETP {\bf 44}, 443 (1976)
  [Zh.\ Eksp.\ Teor.\ Fiz.\  {\bf 71}, 840 (1976)].

E.~A.~Kuraev, L.~N.~Lipatov and V.~S.~Fadin,
  Sov.\ Phys.\ JETP {\bf 45}, 199 (1977)
  [Zh.\ Eksp.\ Teor.\ Fiz.\  {\bf 72}, 377 (1977)].

I.~I.~Balitsky and L.~N.~Lipatov,
  Sov.\ J.\ Nucl.\ Phys.\  {\bf 28}, 822 (1978)
  [Yad.\ Fiz.\  {\bf 28}, 1597 (1978)].

\bibitem{GLR} L.V. Gribov,  E.M. Levin and M.G. Ryskin, Phys. Rept. 100 (1983) 1.

\bibitem{LR}   E.M. Levin and M.G. Ryskin, Phys. Rept. 189 (1990) 267.

\bibitem{PDG} V.A.Khoze, M.G.Ryskin and M. Tasevsky, Chapter 20 in P.~A.~Zyla \textit{et al.} [Particle Data Group],
PTEP \textbf{2020}, no.8, 083C01 (2020).

  \bibitem{GW}
  M.~L.~Good and W.~D.~Walker,
  Phys.\ Rev.\  {\bf 120}, 1855 (1960).

\bibitem{NLL} 
V.~S.~Fadin and L.~N.~Lipatov,
  Phys.\ Lett.\ B {\bf 429}, 127 (1998)
  [hep-ph/9802290];
  
 M.~Ciafaloni and G.~Camici,
  Phys.\ Lett.\ B {\bf 430}, 349 (1998)
  [hep-ph/9803389];\\
  G.~P.~Salam,
  JHEP {\bf 9807}, 019 (1998)
  [hep-ph/9806482];\\
  M.~Ciafaloni and D.~Colferai,
  Phys.\ Lett.\ B {\bf 452}, 372 (1999)
  [hep-ph/9812366].\\
S.J. Brodsky et al., JETP Lett. 70 (1999) 155.

  \bibitem{FC} E.L. Feinberg and D.S. Chernavski, Usp. Fiz. Nauk {\bf 82} (1964) 41.  
\bibitem{Grb} V.N. Gribov, Yad. Fiz. {\bf 9} (1969) 640, Sov.J.Nucl.Phys. {\bf 9} (1969) 369. 
\bibitem{DL} 
A.~Donnachie and P.~V.~Landshoff,
  Nucl.\ Phys.\ B {\bf 231}, 189 (1984).
\bibitem{Burq} J.P. Burg et al., Nucl. Phys. B217 (1983) 285.  
  
  \bibitem{TOT-lM}
TOTEM
collaboration, G. Antchev et al.,
Eur. Phys. Lett. 101 (2013) 21003.
  \bibitem{Em}    E.G. de Oliveira, A.D. Martin and M.G. Ryskin,  Phys.Lett. B 695 (2011) 162-164, arXiv: 1010.1366 [hep-ph].
  \bibitem{Luna}
 E.~G.~S.~Luna, V.~A.~Khoze, A.~D.~Martin and M.~G.~Ryskin,
Eur.\ Phys.\ J.\ C {\bf 59} (2009) 1
[arXiv:0807.4115 [hep-ph]].
\bibitem{elastic}
  TOTEM Collaboration, G. Antchev {\it et al.}, Europhys. Lett. {\bf 101}, 21002 (2013);\\
   TOTEM Collaboration, G. Antchev {\it et al.}, Eur. Phys. J. {\bf C79}, 861 (2019);\\
      TOTEM Collaboration, G. Antchev {\it et al.}, Eur. Phys. J. {\bf C80}, 91 (2020);\\
 UA4 Collaboration, M. Bozzo {\it et al.} ,Phys. Lett. {\bf B147}, 385 (1984);\\
UA4/2 Collaboration, C. Augier {\it et al.}, Phys. Lett. {\bf B316}, 448 (1993); \\
UA1 Collaboration, G. Arnison {\it et al.}, Phys. Lett. {\bf B128}, 336 (1982); \\
E710 Collaboration, N.A. Amos {\it et al.}, Phys. Lett. {\bf B247}, 127 (1990);\\
CDF Collaboration, F. Abe {\it et al.}, Phys. Rev. {\bf D50}, 5518 (1994);\\
N. Kwak et al., Phys. Lett. {\bf B58}, 233 (1975);\\
U. Amaldi et al., Phys. Lett. {\bf B66}, 390 (1977);\\
L. Baksay et al., Nucl. Phys. {\bf B141}, 1 (1978); \\
U. Amaldi et al., Nucl. Phys. {\bf B166}, 301 (1980);\\
M. Bozzo et al., Phys. Lett {\bf B155}, 197 (1985);\\
D0 Collaboration, V.M. Abazov et al., Phys. Rev. {\bf D86}, 012009 (2012).

\bibitem{COMPETE} 

  J.~R.~Cudell {\it et al.} [COMPETE Collaboration],
  Phys.\ Rev.\ Lett.\  {\bf 89}, 201801 (2002)
  [hep-ph/0206172].

  C.~Patrignani {\it et al.} [Particle Data Group],
  Chin.\ Phys.\ C {\bf 40}, no. 10, 100001 (2016), Section 51.

 \bibitem{cdf} 	
Konstantin A. Goulianos, J. Montanha,  Phys.Rev. D59 (1999) 114017 [arXiv: hep-ph/9805496].

\bibitem{ATL} 	
G. Aad et al. (ATLAS Collaboration), JHEP 2002 (2020) 042; 
erratum JHEP10 (2020) 182 
arXiv:1911.00453 [hep-ex].
\bibitem{cms} Vardan Khachatryan (CMS Collaboration),
 Phys. Rev. D 92 (2015) 012003, arXiv: 1503.08689 [hep-ex].
\bibitem{NM127} Paul Newman and Marek Tasevsky, private communication.
 \bibitem{Ryskin:2011qh}
M.~G.~Ryskin, A.~D.~Martin and V.~A.~Khoze,
J. Phys. G \textbf{38}, 085006 (2011)
[arXiv:1105.4987 [hep-ph]].

\bibitem{KMR14} 
V.~A.~Khoze, A.~D.~Martin and M.~G.~Ryskin,  
    Int.J.Mod.Phys. {\bf A 30} (2015) 1542004, arXiv: 1402.2778 [hep-ph].
\bibitem{KMR-slope} 
V.~A.~Khoze, A.~D.~Martin and M.~G.~Ryskin,
  Eur.\ Phys.\ J.\ C {\bf 73}, 2503 (2013)
  [arXiv:1306.2149 [hep-ph]].

\bibitem{GLM} 
E.~Gotsman, E.~Levin and U.~Maor,
  Int.\ J.\ Mod.\ Phys.\ A {\bf 30} (2015) no.08,  1542005
  [arXiv:1403.4531 [hep-ph]].
  
  \bibitem{KKPT} 
  A.~B.~Kaidalov, V.~A.~Khoze, Y.~F.~Pirogov and N.~L.~Ter-Isaakyan,
 Phys.\ Lett.\  {\bf 45B} (1973) 493.

\bibitem{FF}
 R.~D.~Field and G.~C.~Fox,
 Nucl.\ Phys.\ B {\bf 80}, 367 (1974).

\bibitem{KMR11}    M.G. Ryskin, A.D. Martin, V.A. Khoze, Eur.Phys.J.C 71 (2011) 1617;
        1102.2844 [hep-ph].

\end{document}